\documentclass[aps,pre,reprint,superscriptaddress,amsmath]{revtex4-2}
\usepackage{lipsum}
\usepackage{graphicx}
\usepackage{subfigure}
\usepackage{textcomp}
\usepackage{setspace}
\begin{document}
\bibliographystyle{apsrev4-1} 
\title{Pressure-induced structural phase transitions of zirconium:\\ An \emph{ab initio} study based on statistical ensemble theory}
\author{Bo-Yuan Ning}
\email{byning@fudan.edu.cn}
\affiliation{Institute of Modern Physics, Fudan University, Shanghai, 200433, China}
\affiliation{Applied Ion Beam Physics Laboratory, Fudan University, Shanghai, 200433, China}
\affiliation{Department of Materials Science and Engineering,
  Southern University of Science and Technology, Shenzhen 518055, China}
\date{\today}

\begin{abstract}
  The structural phase behaviors of pure zirconium metal under compressions up to $160$ GPa at room temperature are investigated from the perspective of ensemble theory where the partition function is solved by our recently proposed method with \emph{ab initio} precision. The derived Gibbs free energy is employed as the very criterion to determine phase transitions and the calculated transition pressures of the $\alpha\rightarrow\omega\rightarrow\beta$ are $6.93$ and $24.83$ GPa respectively, the former one of which is so far the only theoretical result agreeing with multiple experimental measurements to our best knowledge. The differences between the obtained parameter-free equation of state and those from latest experiments are less than $1.5\%$ in the whole studied pressure range, and particularly, within $0.7\%$ when the applied pressure exceeds over $40$ GPa, the coincidence of which makes us support the argument that the previously observed anharmonicity-driven isostructural phase transition does not exist in the $\beta$-phase even though the thermal effects at room temperature are confirmed to be nontrivial to the phase stability by our quantitative comparisons with the results at $0$K.
\end{abstract}

\maketitle
\section{Introduction}
Investigations on structural phase transitions of condensed matters under high pressure-temperature conditions are not only directly related to understanding the mechanism of how the planets work\cite{earth1,earth2,eos1},
but also significant for their potential applications as novel functional and structural materials\cite{hp1,hp2,hp3}.
As important fuel cladding materials widely used in industrial nuclear reactors\cite{zircaloy1},
zirconium (Zr) and Zr-based alloys have been extensively studied for their high-temperature properties\cite{zircaloy2,zircaloy3,zircaloy4}, and recently,
the structural transformations of pure Zr metal under high pressures attracted broad interests and controversies.
Among various structural phase transitions,
the isostructural one may be the most unique 
because, in terms of the pure-element crystals,
such a phenomenon so far has been solely confirmed in the cerium metal that a $14$\% volume collapse would take place at the transition pressure of $\sim0.8$ GPa at room temperature\cite{decremps2011,lipp2012} and the underlying mechanism remains to be in disputes over the role played by 4$f$ electrons\cite{huang2019,casadei2016}, 4$f$-5$f$ electrons\cite{lanata2013,amadon2015} or the lattice vibrations\cite{jeong2004}.
After the denials of osmium\cite{os} and boron\cite{boron},
experimental observations and theoretical computations revealed a possibility of Zr metal as the second pure-element candidate possessing the feature of isostructural transformation.

Zr metal is in a hexagonal-closed-packed (HCP) structure ($\alpha$-phase) at ambient conditions,
and transforms into another HCP structure with space group P6/mmm ($\omega$-phase) as the pressure is increased up to $5$-$15$ GPa,
and finally ends up into a body-centered-cubic (BCC) structure ($\beta$-phase) as the pressure reaches $30$-$37$ GPa\cite{akahama1991,zhao2005,stavrou2018,pigott2019,anzellini2020,bannon2021,zong2014}.
When the applied pressures is larger than $56$ GPa, 
Akahama et al.\cite{akahama1991} for the first time observed that the $\beta$-phase would undergo an isostructural transition to $\beta'$-phase associated with a discontinued $1.2\%$ volume change,
which was later theoretically supported by Trubitsin et al.\cite{yu2008} analyzing the stability of the longitudinal and transverse phonon modes in the $\beta$-phase.
Despite of the following experiment\cite{zhao2005} and \emph{ab initio} computations based on either $0$K enthalpy\cite{zhang2010} or quasi-harmonic approximated (QHA) lattice dynamics\cite{greeff2005,hao2008,hu2011,wang2011} failing to find the phenomenon,
the experiment by Stavrou et al.\cite{stavrou2018} successfully reproduced the transition of $\beta\rightarrow\beta'$ at the transition pressure of $58$ GPa accompanied with a prominent $4\%$ volume collapse and
attributed the driving force to the anharmonicity of lattice dynamics,
which was manifested by their quantum molecular dynamics (QMD) simulations of a quenching process from $1000$K to $300$K at a cooling rate of $100$K per picosecond.
Gal\cite{gal2021} further indicated a second isostructural transition of $\beta'\rightarrow\beta''$ above $110$ GPa
by more cautiously fitting the data from Ref.\cite{stavrou2018} to two commonly-used empirical Birch-Murnaghan\cite{bm} and Vinet\cite{vinet} equations of states (EOSs).
Nevertheless,
after hydrostatic and non-hydrostatic runs focusing on the $\beta$-phase with the obtained EOSs differing about $3\%$ from each other,
Pigott et al.\cite{pigott2019} failed to observe the claimed volume collapse
and speculated the reason to be the impurities effects on the Zr samples.
At the same time, 
Anzellini et al.\cite{anzellini2020} conducted the compressions in a wider pressure rage up to $150$ GPa and
the results showed no discontinuities where they excluded the impacts from the sample impurity
but regarded the non-hydrostaticity\cite{daniel2005} to be accounting for the observed isostructural transitions in Refs.\cite{akahama1991} and \cite{stavrou2018}.
The latest experiment of the $\beta$-phase by Bannon et al.\cite{bannon2021} also exhibited no evidence of the isostructural phenomenon and the inaccurately calibrated EOS of the pressure markers was considered to be the very cause of the observed abnormal volume collapse in Ref.\cite{stavrou2018}.

In addition to the discrepancies of the isostructural transformation in the $\beta$-phase,
as pointed out by Bannon\cite{bannon2021},
another unsolved problem of Zr is that
the two transition pressures of the $\alpha\rightarrow\omega\rightarrow\beta$ determined by \emph{ab initio} computations\cite{greeff2005,yu2008,hao2008,zhang2010,hu2011,wang2011,schnell2006,hao2008_2},
with or without phonon vibration included,
are much lower than all the experimental observations,
especially the one for the $\alpha\rightarrow\omega$ transition as summarized in Table.\ref{tab1}.
From the perspective of thermodynamics,
free energy (FE) is the very driving force that governs the phase stability, and actually,
the statistical ensemble theory has already paved a rigorous way to derive all the thermodynamic state functions as long as the partition function (PF) is solved\cite{fe,pes1}.
Unfortunately,
the complex high-dimension integral in the PF severely hinders the theory to be applied in the field of condensed matters so that the capability of ensemble theory has been seriously questioned when dealing with the first-order phase transitions of condensed matters\cite{pfconden,mce}.
Despite of great progresses made to solve the PF\cite{ce2020,fe1},
the computational efficiency is still a bottleneck\cite{DOS} that limits current algorithms to afford \emph{ab initio} computations\cite{dft1,dft3,dft5,dft2} to characterize interatomic interactions but resort to empirical potentials for realistic condensed-matter materials,
which leads to large deviations from experimentally determined phase behaviors under high pressure-temperature conditions\cite{nspv}.

We recently put forward a parameter-free direct integral approach (DIA) to the PF of condensed matters\cite{nby},
the ultrahigh efficiency and precision of which were examined by previous comparisons with state-of-the-art sampling algorithm\cite{glc1} and the QHA phonon model\cite{glc2}.
The method has been successfully applied to study the EOS of copper\cite{nby}, the optimum growth condition for two-dimension materials\cite{lyp} and structural phase transitions of vanadium\cite{nby2022v} and aluminum\cite{nby2022al} with the density functional theory (DFT) computations incorporated for the interatomic interactions.
In this work,
we aim at applying DIA to compute the Gibbs FE and EOS of pure Zr metal in order to settle down the discrepancies from a different theoretical path.
The paper is organized as follows:
The theoretical model of DIA, the regarding implementations to the $\alpha$, $\omega$ and $\beta$-Zr,
and the detailed parameters of the DFT computations are presented in Sec.\ref{sec:2}.
The obtained phase transitions and the EOSs by DIA are discussed and compared with both the results at $0$K and experiments in Sec.\ref{sec:3}.
Finally, a conclusion is made in Sec.\ref{sec:4}.

\section{Method}
\label{sec:2}
\begin{table}
\caption{\label{tab1}Transition pressures (units in GPa) for crystalline Zr at $300$K determined by theoretical and  experimental works.} 
\begin{ruledtabular}
\begin{tabular}{cccc}
Theoretical Works & $\alpha\rightarrow\omega$ & $\omega\rightarrow\beta$ & $\beta\rightarrow\beta'$ \\
  \hline
  present work (DIA) & 6.93 & 24.83 & \emph{not found}\\
  present work ($0$K) & 0.51 & 27.86 & /\\
  Trubitsin et al.\cite{yu2008}\footnotemark[1] & 3.99 & 22.65 & 46.08\\
  Zhang et al.\cite{zhang2010}\footnotemark[2] & 0.14 & 27.01 & \emph{not found}\\
  Wang et a.\cite{wang2011}\footnotemark[2] & -3.7 & 32.4 & /\\         
  Schnell et al.\cite{schnell2006}\footnotemark[2] & $< 0$ & 28.2 & /\\
  Hao et al.\cite{hao2008_2}\footnotemark[2] & $< 0$ & 26.8 & /\\
  Greeff\cite{greeff2005}\footnotemark[3] & 2.2 & 32.6 & \emph{not found}\\
  Hao et al.\cite{hao2008}\footnotemark[3] & 1.7 & / & /\\
  Hu et al.\cite{hu2011}\footnotemark[3] & 2.1 & 28.4 & \emph{not found}\\
  \hline
  \hline
  Experimental Works &  & &  \\\hline
  Akahama et al.\cite{akahama1991}\footnotemark[4] & 6.7 & 33 & 56 \\
  Zhao et al.\cite{zhao2005}\footnotemark[4] & 5.5 & / & / \\
  Stavrou et al.\cite{stavrou2018}\footnotemark[4] & 12.7 & 30 & 58 \\
  Pigott et al.\cite{pigott2019}\footnotemark[4] & / & 37 & \emph{not found}\\
  Anzellini et al.\cite{anzellini2020}\footnotemark[4] & 14\footnotemark[6] & 35
 & \emph{not found}\\
 & 20\footnotemark[6] & 35 & \\
  Bannon et al.\cite{bannon2021}\footnotemark[4] & 10.7\footnotemark[7] & 34.9\footnotemark[7] & \emph{not found}\\
  & 10.8\footnotemark[7] & 35.0\footnotemark[7] & \\
  & 12.7\footnotemark[7] & 34.6\footnotemark[7] & \\
  Liu et al.\cite{liu2008,liu2007}\footnotemark[5] & 6.8 & / & /\\
\end{tabular}
\end{ruledtabular}
\footnotetext[1]{\emph{Ab initio} phonon mode analysis.}
\footnotetext[2]{\emph{Ab initio} $0$K-enthalpy.}
\footnotetext[3]{\emph{Ab initio} + QHA phonon computations.}
\footnotetext[4]{Static compression + X-ray diffraction.}
\footnotetext[5]{Static compression + Ultrasonic measurement.}
\footnotetext[6]{Values determined in two separate runs.}
\footnotetext[7]{Values determined in three separate runs.}
\end{table}

\begin{figure}
  \centering
  \includegraphics[width=2.5in,height=4.2in]{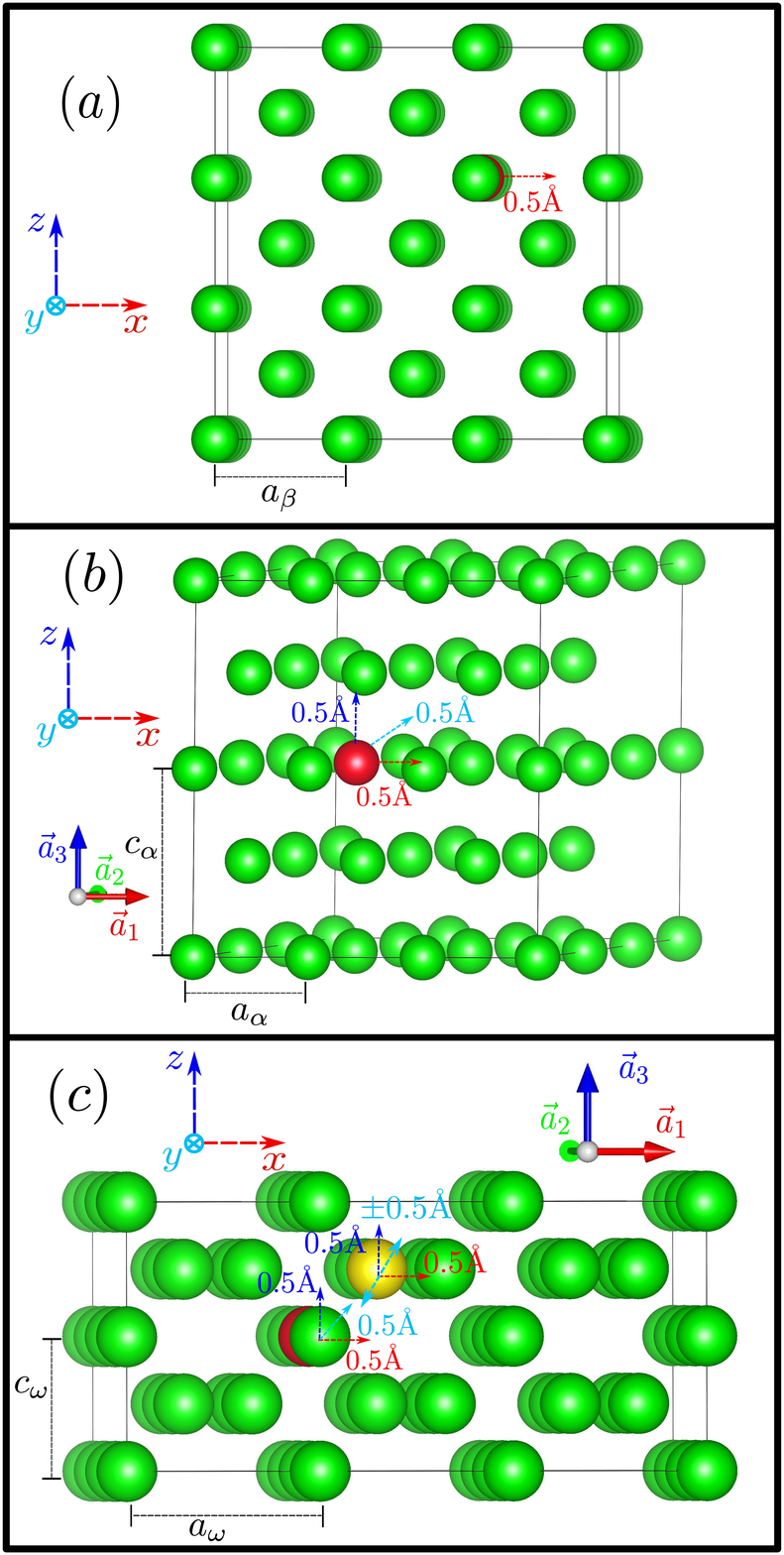}
  \caption{\label{fig:1}(Color Online) Schematic diagrams of implementations of DIA. (a) The $\beta$-phase are placed in a cubic lattice with lattice parameter of $a_{\beta}$, and an arbitrarily selected atom is moved from its lattice site (colored in red) along the $x$ axis by $0.5$ {\AA} (denoted by red arrow) for the computation of effective length $\mathcal L$ defined in Eq.(\ref{eq:6}). (b) The $\alpha$-phase is placed in a hexagonal lattice with lattice parameter of $a_{\alpha}$ and axial ratio $c_{\alpha}/a_{\alpha}$, and an atom is moved from its lattice site (colored in red) along the $x$, $y$ and $z$ axis separately by $0.5$ {\AA} (denoted by red, cyan and blue arrows respectively) for the computations of effective lengths $\mathcal L_x$, $\mathcal L_z$ and $\mathcal L_z$ defined in Eq.(\ref{eq:7}). (c) The $\omega$-phase is placed in a hexagonal lattice with lattice parameter of $a_{\omega}$ and axial ratio $c_{\omega}/a_{\omega}$, and atoms are divided into two groups to apply Eq.(\ref{eq:8}). In the first group, one atom (colored in red) is moved from its lattice site along the $x$, $y$ and $z$ axis separately by $0.5$ {\AA}, and one atom in the second group (colored in yellow) is conducted in the same way except that, along the $y$ axis, the atom is moved in both positive and negative directions by $0.5$ {\AA}. The view angles of the above shown lattices are slightly tilted for a more clear illustration.}
\end{figure}

\subsection{Theoretical Model: DIA}
\label{sec:2:1}
According to ensemble theory,
the PF for a system consists of $N$ particles with their Cartesian coordinate $\mathbf{q}^N=\{\mathbf q_1,\mathbf q_2,\ldots\mathbf q_N\}$ confined within a volume $V$ at temperature $T$ is formulated as
\begin{eqnarray}
  \label{eq:1}
  \mathcal{Z}&=&\frac{1}{N!}\left(\frac{2\pi m}{\beta h^2}\right)^{\frac{3}{2}N}\int d\textbf{q}^N\exp[-\beta U(\textbf{q}^N)] \nonumber \\
             &=&\frac{1}{N!}\left(\frac{2\pi m}{\beta h^2}\right)^{\frac{3}{2}N}\mathcal Q,
\end{eqnarray}
where $h$ denotes the Planck constant,
$m$ the particle mass, $\beta=1/k_BT$ with $k_B$ the Boltzmann constant,
$U(\textbf{q}^N)$ the total potential energy,
and $\mathcal Q=\int d\textbf{q}^N\exp[-\beta U(\textbf{q}^N)]$ the so-called configurational integral (CI)
that is related to the structures of the system at given conditions.
If the CI is solved,
then the pressure-volume ($P$-$V$) EOS and the Gibbs FE ($\mathcal G$) can be computed as
\begin{eqnarray}
  \label{eq:2}
  P&=&\frac{1}{\beta}\frac{\partial\ln\mathcal{Q}}{\partial V}, \\
  \label{eq:3}
  \mathcal G&=&-\frac{1}{\beta}\ln[\frac{1}{N!}\left(\frac{2\pi m}{\beta h^2}\right)^{\frac{3}{2}N}]-\frac{1}{\beta}\ln\mathcal{Q}+PV.
\end{eqnarray}
For a crystalline system with atoms locating on the lattice sites, $\mathbf Q^N$,
and with the total potential energy, $U_0(\mathbf{Q}^N)$,
the model of DIA\cite{nby} firstly introduces transformations as
\begin{equation}
  \label{eq:4}
  \mathbf q'^N=\mathbf q^N-\mathbf Q^N,\ U'(\mathbf q'^N)=U(\mathbf q'^N)-U_0(\mathbf Q^N), 
\end{equation}
where $\mathbf q'^N$ represents the displacements of atoms away from their lattice sites
and $U'(\mathbf q'^N)$ stands for the corresponding differences of total potential energy with respect to the $U_0(\mathbf Q^N)$.
Therefore,
the CI can be expressed as
\begin{equation}
  \label{eq:4a}
  \mathcal Q=e^{-\beta U_0(\mathbf{Q}^N)}\int e^{-\beta U'(\mathbf q'^N)}d\mathbf q'^N.
\end{equation}
Based on our reinterpretations of integrals,
the $3N$-fold integral in Eq.(\ref{eq:4a}) is mapped to an effective $3N$-dimension volume,
and may be further simplified as
\begin{eqnarray}
  \label{eq:5}
  \mathcal{Q}&=&e^{-\beta U_0(\mathbf Q^N)}\times \nonumber \\
  &&\prod_{i=1}^{N}\int e^{-\beta U'(q'_{i_x})}dq'_{i_{x}}\int e^{-\beta U'(q'_{i_y})}dq'_{i_{y}}\int e^{-\beta U'(q'_{i_z})}dq'_{i_{z}} \nonumber \\
  &=&e^{-\beta U_0(\mathbf Q^N)}\prod_{i=1}^{N}\mathcal L_{i_{x}}\mathcal L_{i_{y}}\mathcal L_{i_{z}},
\end{eqnarray}
where $q'_{i_{x(y,z)}}$ denotes the distance of the $i$th atom moving along the $x$ (or $y$, $z$) axis relative to its lattice site while the other two degrees of freedom of the atom and all the other atoms are kept fixed,
$U'(q'_{i_{x(,y,z)}})$ represents the potential-energy curve (function) felt by the moved atom,
and $\mathcal L_{i_{x(y,z)}}$ is called the effective length of the atom along the $x$ (or $y$, $z$) axis.
It should be emphasized here that Eq.(\ref{eq:5}) is the only approximation in the model of DIA and the original physics picture as well as detailed mathematical proof is referred to Ref.\cite{nby}.

For pure-element crystals with BCC structure
where all the atoms in the lattice are geometrically equivalent and
the potential-energy curve $U'_x$ felt by an arbitrary atom moving along $x$ axis is the same as the one
along $y$ or $z$ axis,
the CI can thus be simplified as
\begin{equation}
  \label{eq:6}
  \mathcal Q=e^{-\beta U_0(\mathbf Q^N)}\mathcal L^{3N},
\end{equation}
where $\mathcal L$ represents the effective length of an arbitrary atom along either $x$, $y$ or $z$ axis. 
For the pure-element crystals with HCP structure
where the atoms are geometrically equivalent but the three effective lengths $\mathcal L_{x(y,z)}$ of each atom are not equivalent to each other, 
the CI consequently turns into
\begin{equation}
  \label{eq:7}
  \mathcal Q=e^{-\beta U_0(\mathbf Q^N)}\left(\mathcal L_x\mathcal L_y\mathcal L_z\right)^N.
\end{equation}
For crystals with more complex structures or with multiple elements,
the atoms may be divided into $M$ groups with each containing a number of $N_I$ equivalent atoms
on the basis of the geometric feature of the lattice,
and the CI would be
\begin{equation}
  \label{eq:8}
  \mathcal Q=e^{-\beta U_0(\mathbf Q^N)}\prod_{I=1}^{M}\left(\mathcal L_{I_{x}}\mathcal L_{I_{y}}\mathcal L_{I_{z}}\right)^{N_I},
\end{equation}
the detailed applications of which can be found in our previous works of large molecules\cite{nby} and two-dimension materials\cite{lyp}.

\subsection{Implementation of DIA to Zr}
\label{sec:2:2}
As shown in Fig.\ref{fig:1}(a),
the $\beta$-phase is placed in a $3\times3\times3$ cubic supercell with two atoms in the unit cell
where the lattice vectors are set as $\mathbf{a}_1=a_{\beta}(1, 0, 0)$, $\mathbf{a}_2=a_{\beta}(0, 1, 0)$ and $\mathbf{a}_3=a_{\beta}(0, 0, 1)$ with $a_{\beta}$ being the lattice parameter,
and the basis vectors are $\mathbf b_1=(0, 0, 0)$ and $\mathbf b_2=(\frac{1}{2}, \frac{1}{2}, \frac{1}{2})$.
The DIA is applied by using Eq.(\ref{eq:6}) and
 an arbitrary atom is selected to move $0.5$ {\AA} from the lattice site at a step of $0.05$ {\AA} long its $x$ axis to obtain the corresponding potential-energy curve.

For the $\alpha$-phase, as shown in Fig.\ref{fig:1}(b),
the atoms are placed in a $3\times3\times2$ hexagonal supercell with two atoms in the unit cell
where the lattice vectors are $\mathbf{a}_1=a_{\alpha}(1, 0, 0)$, $\mathbf{a}_2=a_{\alpha}(\frac{1}{2}, \frac{\sqrt 3}{2}, 0)$ and $\mathbf{a}_3=a_{\alpha}(0, 0, \frac{c_{\alpha}}{a_{\alpha}})$ with $a_{\alpha}$ being the lattice parameter and $\frac{c_{\alpha}}{a_{\alpha}}$ the axial ratio,
and the basis vectors are $\mathbf b_1=(0, 0, 0)$ and $\mathbf b_2=(\frac{1}{3},\frac{1}{3},\frac{1}{2})$.
The DIA is applied by using Eq.(\ref{eq:7})
and an arbitrarily selected atom is moved  from the lattice site along the $x$, $y$ and $z$ axes separately by $0.5$ {\AA} at a step of $0.05$ {\AA} to obtain the three potential-energy curves.

The $\omega$-phase has the most complex structure among the three phases and a detailed elaboration of the structure can be found in supplementary materials.
As shown in Fig.\ref{fig:1}(c),
the atoms are placed in a $3\times3\times2$ hexagonal supercell with three atoms in the unit cell
where the lattice vectors are set as $\mathbf{a}_1=a_{\omega}(1, 0, 0)$, $\mathbf{a}_2=a_{\omega}(-\frac{1}{2}, \frac{\sqrt 3}{2}, 0)$ and $\mathbf{a}_3=a_{\omega}(0, 0, \frac{c_{\omega}}{a_{\omega}})$ with $a_{\omega}$ being the lattice parameter and $\frac{c_{\omega}}{a_{\omega}}$ the axial ratio,
and the basis vectors are $\mathbf b_1=(0, 0, 0)$, $\mathbf b_2=(\frac{1}{3},\frac{2}{3},\frac{1}{2})$,
$\mathbf b_3=(\frac{2}{3},\frac{1}{3},\frac{1}{2})$.
Since neither the geometric symmetry of all the atoms nor the three potential-energy curves of each atom are equivalent,
to apply DIA to the $\omega$-structure,
Eq.(\ref{eq:8}) has to be used and the atoms are divided into two groups that 
the first one includes the atoms in the bottom layer (and equivalent repeated layers)
and the second group includes the atoms in layers with $z=\frac{1}{2}c_\omega$ (and equivalent repeated layers).
In the first group,
the actions on an arbitrarily selected atom are the same as those on the $\alpha$-structure,
and in the second group,
the arbitrarily selected atom is moved along its $x$ and $z$ axes separately by $0.5$ {\AA} for the potential-energy curves,
while, for obtaining the curve along the $y$ axis,
the atom is moved $0.5$ {\AA} in both positive and negative directions due to the asymmetric geometry of the $\omega$-structure (see details in supplementary materials).
For all the three structures,
the potential energies at each moving step are calculated by the DFT computations and
afterwards smoothened by the spline interpolation algorithm\cite{spl1,spl2} (see potential-energy curves displayed in the supplementary materials).

\subsection{Details of DFT Computations}
\label{sec:2:3}
The DFT computations are all conducted in the Vienna Ab initio Simulation Package\cite{VASP1,VASP2} where the projector-augmented wave formalism\cite{paw1,paw2} is used for the pseudopotential and the general gradient approximation of the Perdew-Burke-Ernzerhf parametrizations\cite{pbe} is adopted for the exchange-correlation functional with 12 valence electrons ($4s^24p^65s^24d^2$) considered.
The Gaussian smearing method with a smearing energy of $0.026$ eV is applied in order to smooth the transition of the electron number of orbital occupation\cite{liu2021},
together with $\Gamma$-centered $7\times7\times7$, $5\times5\times11$ and $9\times9\times9$ uniform $k$-mesh grids being set to sample the Brillouin zone of the $\alpha$, $\omega$ and $\beta$-phases by the Monkhorst-Pack scheme\cite{monkhorst} respectively,
$400$ eV as the cut-off energy of the plane-wave basis,
and $1\times10^{-6}$ eV as the stop condition for the electron self-consistent calculations of the total energy.

\section{Results and Discussions}
\label{sec:3}
\subsection{Phase Transition of $\alpha\rightarrow\omega$ at $0$K and $300$K}
\label{sec:3:1}
\begin{figure}
  \centering
  \includegraphics[width=3.1in,height=2.3in]{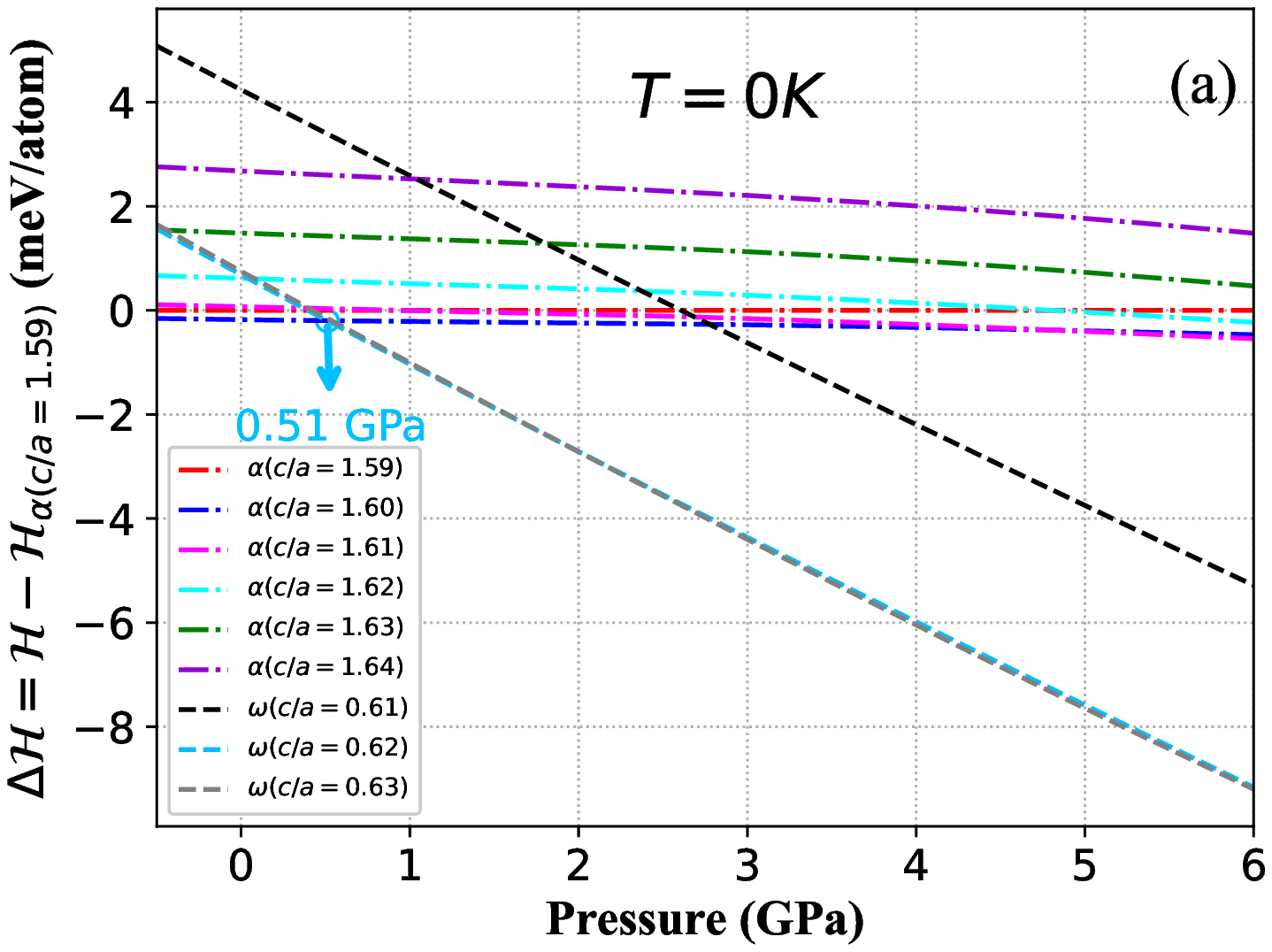}
  \includegraphics[width=3.1in,height=2.3in]{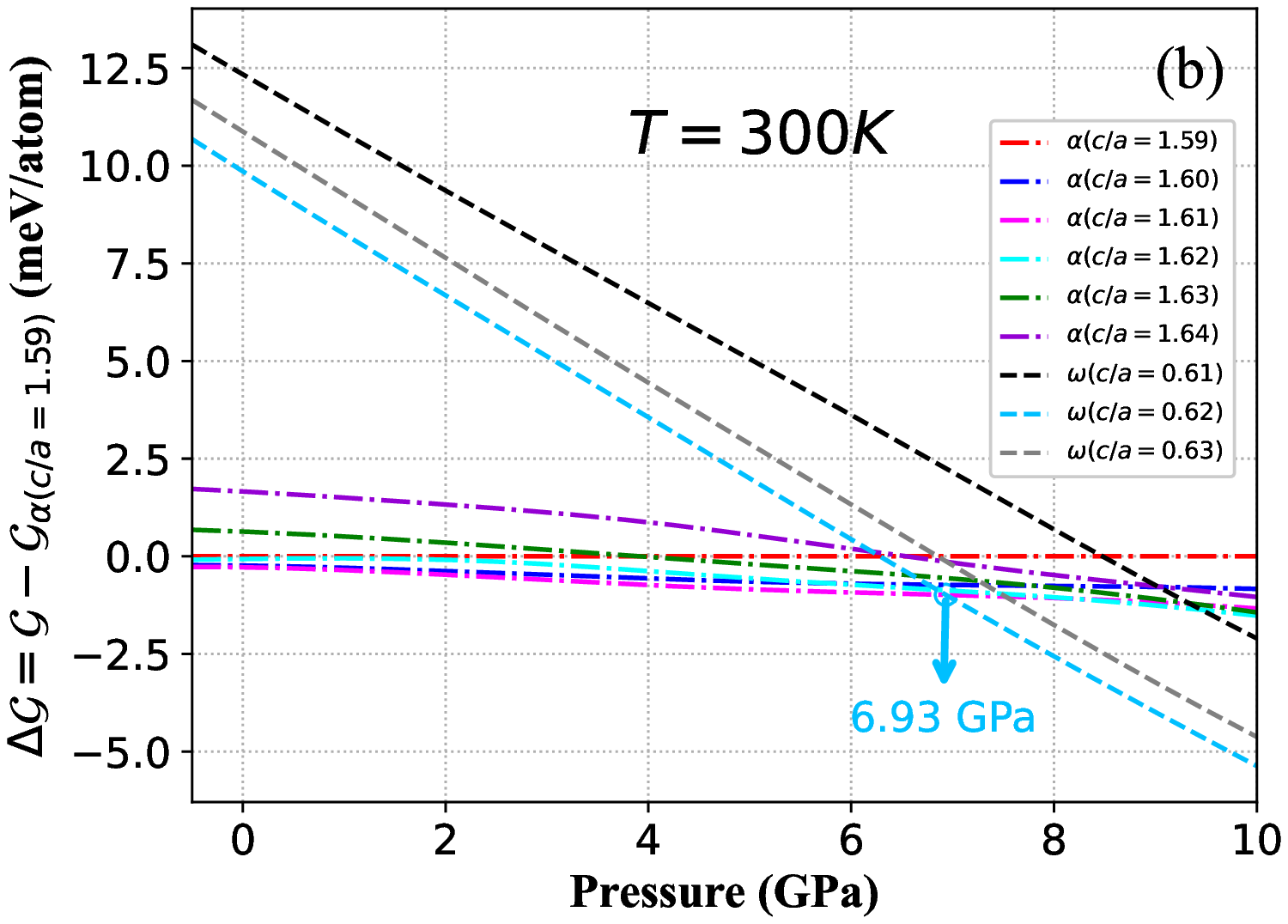}
  \caption{\label{fig:2}(Color Online) The $\alpha\rightarrow\omega$ phase transition.
    (a) The enthalpies at $0$K of five $\alpha$-structures with axial ratio from $1.59$-$1.64$ (denoted by blue, magenta, cyan, green and violet dashed-dotted lines respectively) and of three $\omega$-structures with axial ratio from $0.61$-$0.63$ (denoted by black, skyblue and grey dashed lines respectively) relative to that of the $\alpha$-structure with axial ratio of $1.59$ (red dashed-dotted line). (b) Similar to (a) except that the comparisons are on the basis of Gibbs FEs at $300$K calculated by DIA.}
\end{figure}

We first consider the $\alpha\rightarrow\omega$ phase transition at $0$K, 
where the Gibbs FE becomes equivalent to enthalpy as $\mathcal H=U_0+P_{0}V$ with $U_0$ the total potential energy and $P_{0}=-\partial U_0/\partial V$ the pressure at $0$K.
Although the ideal axial ratios of $\alpha$ and $\omega$-structures are $c_\alpha/a_\alpha=\sqrt{8/3}$ and $c_\omega/a_\omega=\sqrt{3/8}$ respectively,
the realistic values of the most stable structures may vary from the ideal ones, and as a result,
six $\alpha$-structures with axial ratio from $1.59$-$1.64$ and three $\omega$-structures with axial ratio from $0.61$-$0.63$ are considered, 
the calculated enthalpies of which with respect to the $\alpha$-structure with axial ratio of $1.59$ are shown in Fig.\ref{fig:2}(a).
By the identifications of the one with minimum enthalpy, 
the axial ratio of the stable $\alpha$-structure remains to be $1.60$ from $0$ GPa up to $\sim 5$ GPa and then turns to be $1.61$ afterwards.
As to the $\omega$-phase,
the entropy differences between the structure with axial ratio of $0.62$ and that of $0.63$ are quite negligible from $0$-$6$ GPa,
while the entropy of the one with $0.61$ is over $3$ meV/atom larger than the other two. 
The axial ratio of stable $\omega$-structure is determined to be $0.62$ from $0$ GPa to $\sim2.5$ GPa and then to be $0.63$ till $6$ GPa.

With the determined stable $\alpha$ and $\omega$-structures versus the changes of pressures,
the phase transition of $\alpha\rightarrow\omega$ at $0$K is identified at the cross point between the relative enthalpy curves belonged to the two stable structures shown in Fig.\ref{fig:2}(a) and
the transition pressure locates at $0.51$ GPa,
which, as can be seen in Table.\ref{tab1},
agrees with the $0.14$ GPa from Ref.\cite{zhang2010} where the same frozen-core pseudopotential DFT method was used.
In spite of the disagreement of the negative values from Refs.\cite{wang2011,schnell2006,hao2008_2},
it is notable that the calculated transition pressures at $0$K,
though slightly differ from each other,
are all close to $0$ GPa and quite different from the experimentally measured pressure range $5$-$15$ GPa,
which indicates the potential influences from the neglected thermal contributions on the phase transition.

\begin{figure}
  \centering
  \includegraphics[width=3.1in,height=2.3in]{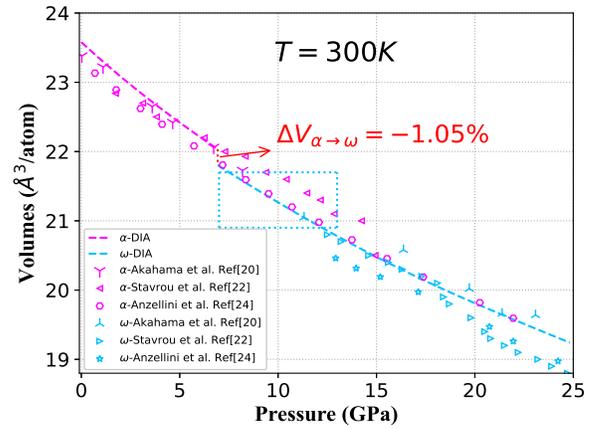}
  \caption{\label{fig:3}(Color Online) $P$-$V$ EOSs of the $\alpha$-phase (magenta dashed line) and the $\omega$-phase (skyblue dashed line) at $300$K by DIA along with those from experiments in a pressure range of $0$-$25$ GPa. The region encircled by the skyblue rectangle is inferred to a possible two-phase coexistence for experimental observations.}
\end{figure}

The Gibbs FEs at $300$K of the above considered $\alpha$ and $\omega$-structures are computed by DIA using Eqs.(\ref{eq:7}) and (\ref{eq:8}) respectively,
and the corresponding differences relative to the $\alpha$-structure with axial ratio of $1.59$ are shown in Fig.\ref{fig:2}(b).
Different from the case at $0$K, 
it is found that the axial ratio of stable $\alpha$-structure at $300$K turns to be $1.61$ from $0$-$8$ GPa and becomes $1.62$ till $10$ GPa.
As to the $\omega$-phase,
the differences of the Gibbs FE are getting prominently larger to be $\sim1$ meV/atom between the structures with axial ratio of $0.62$ and of $0.63$, and
unambiguously identify the former one to be the stable $\omega$-structure as the pressure is increased up to $10$ GPa.

After the confirmations of the stable structures of the two phases,
the transition pressure of $\alpha\rightarrow\omega$ at $300$K is determined to be 6.93 GPa as the cross point highlighted in Fig.\ref{fig:2}(b),
which greatly lifts the pressure value by about $6.5$ GPa compared with the one at $0$K and is by far the only theoretical value coinciding with the experimental pressure range of $5$-$15$ GPa\cite{akahama1991,zhao2005,stavrou2018,anzellini2020,bannon2021,liu2008,liu2007}.
As noticed by Velisavljevic et al.\cite{velisav2011} and Bannon et al.\cite{bannon2021},
experimentally speaking,
the transition pressure is closely related to the purity of Zr sample in a way that the value would be increased as the purity decreases,
which may account for the deviations of our obtained pressure from the reported $12.7$ GPa in Ref\cite{stavrou2018} with $99.5\%$ pure sample or $14$-$20$ GPa in Ref.\cite{anzellini2020} with $98.8\sim99.2\%$ pure samples.
As a comparison,
the observed values of the sample with $99.8\%$ purity in Ref.\cite{akahama1991} and with $99.9995\%$ in Refs.\cite{liu2008,liu2007} are $6.7$ and $6.8$ GPa respectively,
which are in excellent agreement with our result.
The impurity effect on the structural transformations of Zr metal is worthy being carefully investigated in the future but beyond the scope of this work.

According to the determined transition pressure,
the $P$-$V$ EOSs of $\alpha$ and $\omega$-phases in a pressure range of $0$-$25$ GPa at $300$K by Eq.(\ref{eq:2}) are plotted in Fig.\ref{fig:3}  (the higher end of the pressure range of the $\omega$-phase is discussed in the next subsection).
The atomic volume at ambient conditions, $V_0$, by the EOS is determined to be $23.58$ {\AA}$^3$/atom,
and the relative deviations, $|V_{0(DIA)}-V_{0(EXP)}|/V_{0(EXP)}$,
are about $1.11\%$ compared with $23.32$ {\AA}$^3$/atom from Ref.\cite{liu2008} using third-order Birch-Murnaghan EOS,
$1.25\%$ compared with $23.20$ {\AA}$^3$/atom from Ref.\cite{anzellini2020} using Rydberg-Vinet EOS\cite{vinet1989}
and $1.64\%$ compared with $23.2$ {\AA}$^3$/atom from Ref.\cite{gal2021} using Vinet EOS to fit the data measured in Ref.\cite{stavrou2018}.
The relative deviations of $P$-$V$ EOS, $|P_{DIA}-P_{EXP}|/P_{EXP}$,
of the $\alpha$-phase are $0.34\%$, $0.69\%$ and $1.01\%$ compared with the experimental data from Refs.\cite{akahama1991}, \cite{stavrou2018} and \cite{anzellini2020} respectively,
and are $0.68\%$, $1.14\%$ and $1.46\%$ respectively in terms of the $\omega$-phase.
At the transition point of $\alpha\rightarrow\omega$ where $V/V_0=0.93$,
a discontinuity of the EOS accompanies with a $-1.05\%$ volume change,
which is qualitatively consistent with the $-1.26\%$ reported in Ref.\cite{zhao2005}, $-1.36\%$ in Ref.\cite{liu2008} and $-1.5\%$ in Ref.\cite{anzellini2020}.

By the above comparisons,
the computations of Gibbs FE by DIA manifest the importance of thermal contributions to the phase stability of the $\alpha$ and $\omega$-phases at a room-temperature condition,
the phenomenon of which can be also seen in our previous work of aluminum metal under ultrahigh pressures\cite{nby2022al}.
Moreover,
the excellent agreements between our obtained EOSs and those from experiments not only further validate the accuracy of the PF calculated by DIA for the two phases,
but also ensures the accuracy of the following investigations on phase transitions related to $\omega$-phase in higher pressure zone.

\begin{figure}
  \centering
  \includegraphics[width=3.1in,height=2.3in]{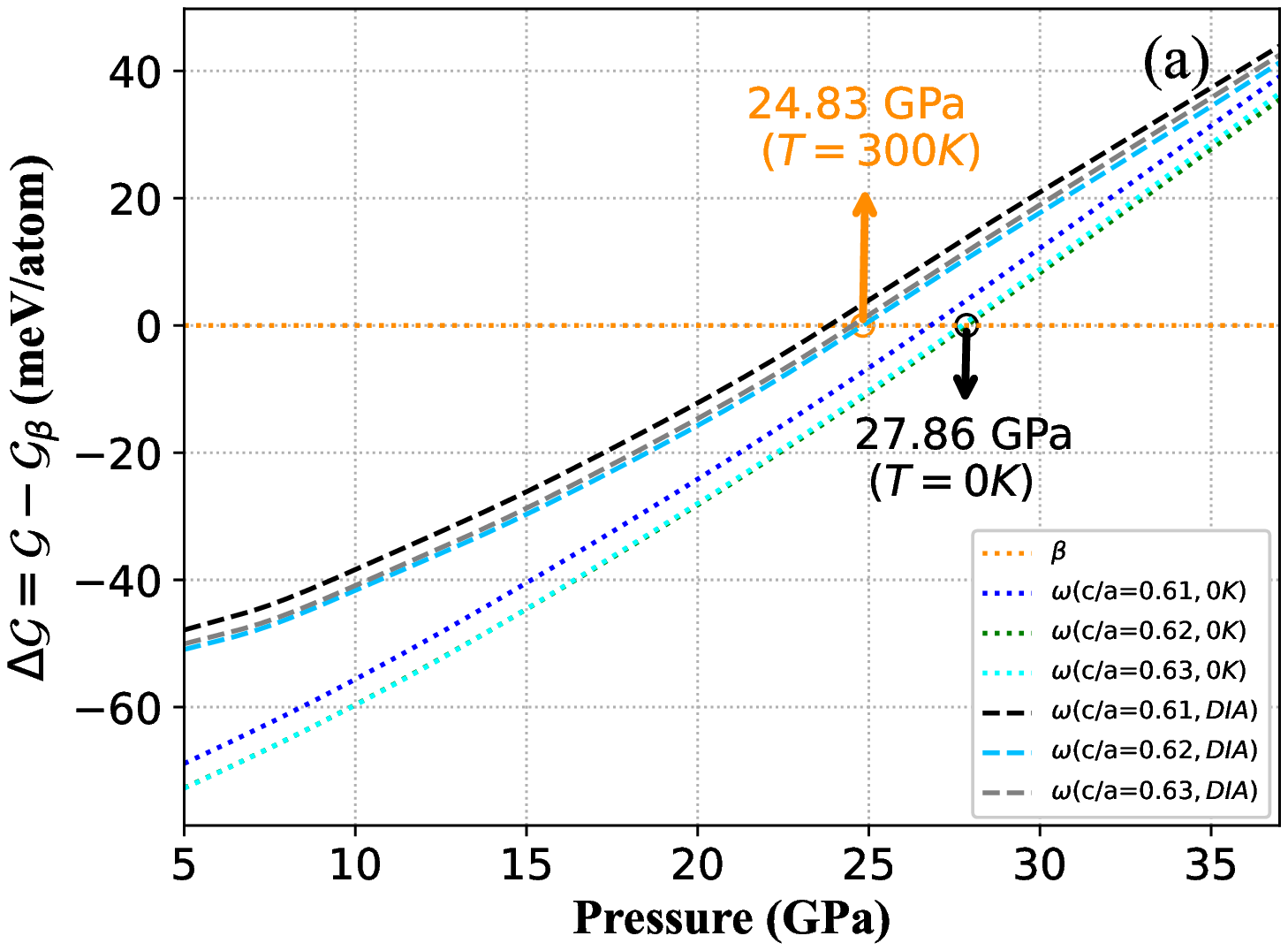}
  \includegraphics[width=3.1in,height=2.3in]{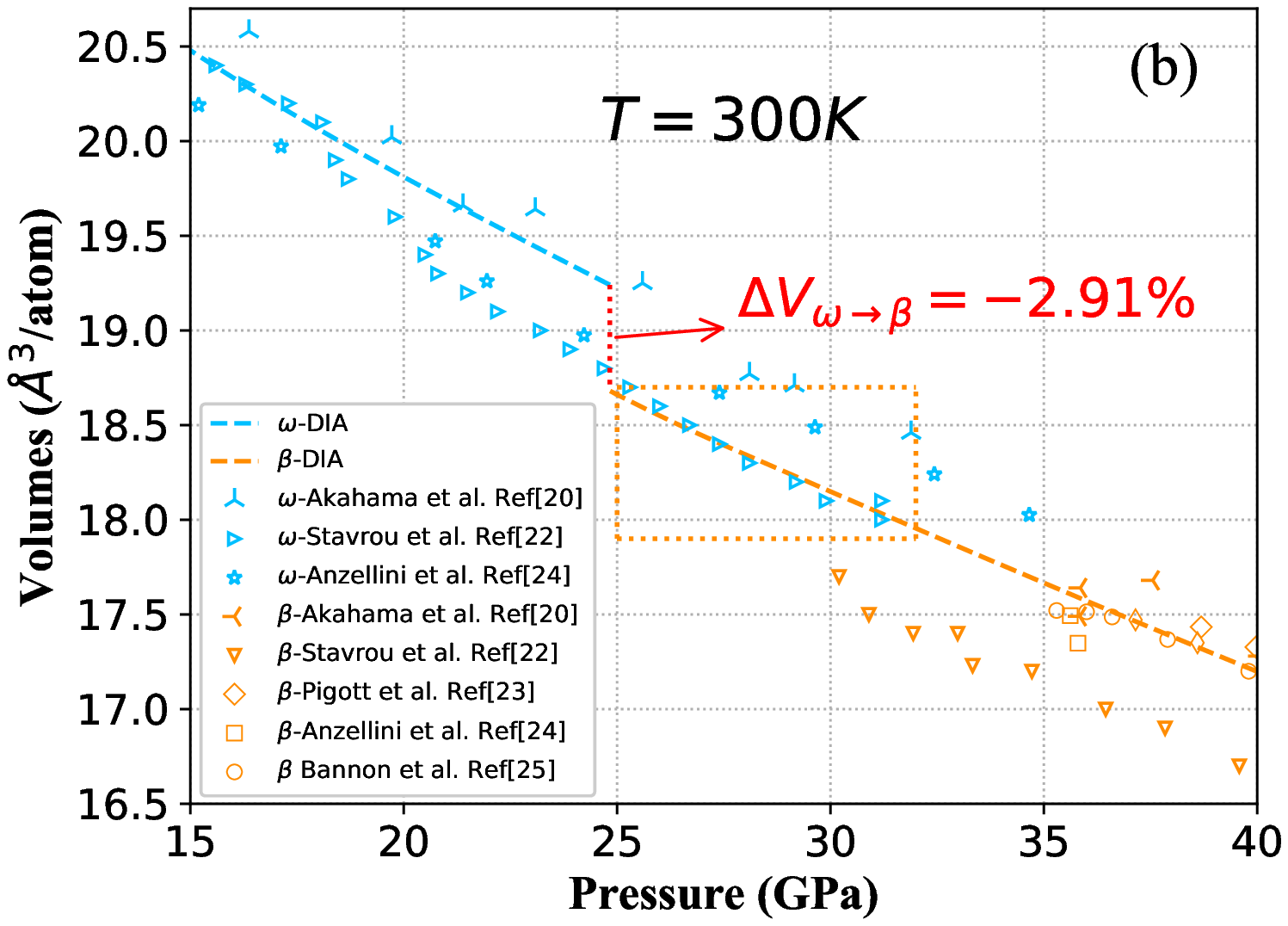}
  \caption{\label{fig:4}(Color Online) (a) The $\omega\rightarrow\beta$ phase transition. The Gibbs FEs at $300$K of three $\omega$-structures with axial ratio of $0.61$, $0.62$ and $0.63$ (denoted in black, skyblue and grey dashed lines respectively) relative to that of $\beta$-structure (denoted in yellow dotted line) and the enthalpy at $0$K of the three structures (denoted in blue, green and cyan dashed lines) relative to that of $\beta$-structure (denoted in yellow dotted line). (b) $P$-$V$ EOSs of $\omega$-phase (skyblue dashed line) and $\beta$-phase (yellow dashed line) at $300$K by DIA along with those from experiments in a pressure range of $15$-$40$ GPa.  The region encircled by the yellow rectangle is inferred to a possible two-phase coexistence for experimental observations.}
\end{figure}

\subsection{Phase Transition of $\omega\rightarrow\beta$ at $0$K and $300$K}
\label{sec:3:2}
For the phase stability in the pressure zone from $5$ to $40$ GPa,
we do not take the $\alpha$-phase into account
because of either the enthalpies or the Gibbs FE of the $\alpha$-structures being apparently larger than that of the $\omega$-structure with the increased pressure according to the trends demonstrated in Figs.\ref{fig:2}(a) and (b).
We still consider the phase transitions at $0$K at first and
the enthalpy differences of the three $\omega$-structures relative to that of $\beta$-structure are shown in the dotted lines in Fig.\ref{fig:4}(a),
where the $\omega$-structure with axial ratio of $0.63$ possesses the minimum enthalpy up to $\sim16$ GPa and then is replaced by the one of $0.62$ till 35 GPa.
The transition pressure identified by the cross point of enthalpy curves between the $\omega$ and $\beta$-structures locates at $27.86$ GPa,
which agrees with previous $0$K-based theoretical predictions about $27$-$28$ GPa from Refs.\cite{zhang2010,schnell2006,hao2008_2} but is lower than the $32.4$ GPa from Ref.\cite{wang2011} that used same pseudopotential method and electronic exchange-correlation functional but considered less valence electrons (4$d^35s^1$) than this work.

The Gibbs FEs at $300$K of the $\omega$ and $\beta$-structures are computed by DIA according to Eqs.(\ref{eq:8}) and (\ref{eq:6}) respectively,
and the differences of the three $\omega$-structures relative to the $\beta$-structure are plotted in the dashed lines in Fig.\ref{fig:4}(a),
which clearly exhibits that the one with axial ratio of $0.62$ continues to be the most stable up to $35$ GPa.
With the determined cross point as highlighted in Fig.\ref{fig:4}(a),
it is interesting to find that the transition pressure at $300$K turns to be $24.83$ GPa,
which shows that, on the contrary to the situation of the $\alpha\rightarrow\omega$,
the thermal contributions reduce the transition pressure of $\omega\rightarrow\beta$ by about $3$ GPa
and lead to more deviations from existing experimental values of $30$-$35$ GPa (listed in Table.\ref{tab1}) when compared with the transition pressure at $0$K.

\begin{figure}
  \centering
  \includegraphics[width=3.4in,height=2.6in]{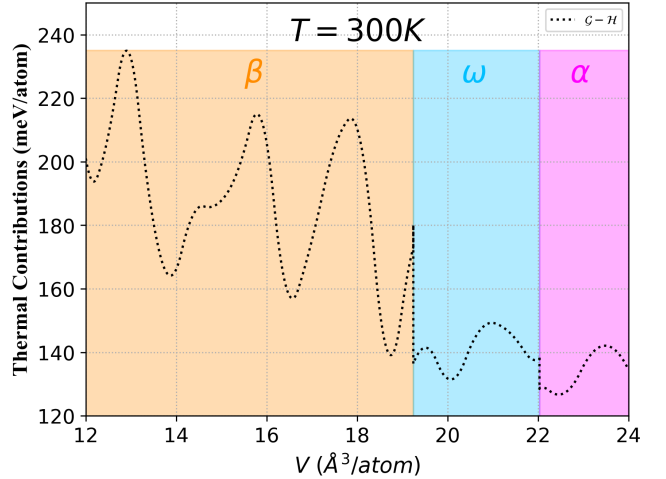}
  \caption{\label{fig:6}(Color Online)  The difference between Gibbs FE at $300$K by DIA and entalpy at $0$K.}
\end{figure}

On the experimental side, 
one possible cause may be from the effects of sample impurity as discussed in the last subsection and 
our obtained pressure is indeed closer to the $33$ and $30$ GPa measured in Refs.\cite{akahama1991,stavrou2018} whose purity is higher than $99.95\%$ while such effects cannot explain why the pressure at $0$K deviates less from experiments.
On the theoretical side,
since there is no adjustable parameters in the model of DIA,
another possible cause may be resulted from the intrinsic precision limits of DFT computations of potential energies,
which has been analyzed in our previous work of aluminum\cite{nby2022al}.
If there exists a $\sim1$ meV/atom error of potential-energy computations,
we estimate a corresponding $4$ meV/atom variation of Helmholtz FE ($\mathcal F$),
which can produce the pressure as $P=-\partial\mathcal F/\partial V$.
According to our computations of $\mathcal F$ and obtained EOSs as shown below,
a $4$ meV/atom variation would lead to a pressure variation of $\Delta P\sim0.1$-$0.2$ GPa
and the total variation of Gibbs FE, $\Delta\mathcal G=\Delta\mathcal F+\Delta P\cdot V$, corresponds to $\sim20$ meV/atom at the transition region,
which is much smaller than the total thermal contributions, $\mathcal G-\mathcal H$,
up to $180$ meV/atom at the transition point (see Fig.\ref{fig:6} in the next subsection).
As a result,
the computational error may limitedly affect the obtained transition pressure but still cannot account for the $5$-$10$ GPa deviations.

We further compare the $P$-$V$ EOS in a pressure range of $15$-$40$ GPa by DIA with recent experiments as shown in Fig.\ref{fig:4}(b),
and the agreement is excellent,
where the relative differences of the $\omega$-phase are $0.86\%$, $1.34\%$ and $1.40\%$ from those in Refs.\cite{akahama1991}, \cite{stavrou2018},
\cite{anzellini2020} respectively,
and of the $\beta$-phase are less than $0.78\%$ compared with Refs.\cite{akahama1991,pigott2019,anzellini2020,bannon2021}.
At the transition of point where $V/V_0=0.82$,
the volume change is determined to be $-2.9\%$ and coincides with the measured $-2.0\%$ in Ref.\cite{anzellini2020}.
In the pressure range of $25$-$32$ GPa,
it is noticeable that the calculated EOS of $\beta$-phase (encircled by a yellow rectangle in Fig.\ref{fig:4}(b)) perfectly agrees with the observed data identified as $\omega$-phase in Ref.\cite{stavrou2018} with a relative difference of $0.23\%$,
as well as of $1.54\%$ compared with the data in Ref.\cite{anzellini2020}, and for the $\alpha\rightarrow\omega$ transition,
a similar phenomenon also exists in the pressure range of $8$-$12$ GPa
where the calculated EOS of $\omega$-phase (encircled by a skyblue rectangle in Fig.\ref{fig:3}) coincides perfectly with the data of $\alpha$-phase in Ref.\cite{anzellini2020}.
Since there are affirmative observations of $\alpha$ and $\omega$-phase coexistence in the high purity Zr sample\cite{liu2007},
and our former analysis excludes large influences from either purity effects or computational variations,
here we suggest that the onset of the $\omega\rightarrow\beta$ transition is in advance than those reported in experiments,
which may be more difficult to be experimentally discerned due to a two-phase coexistence
and needs to be scrutinized in future experiments.

\begin{figure}
  \centering
  \includegraphics[width=3.4in,height=2.6in]{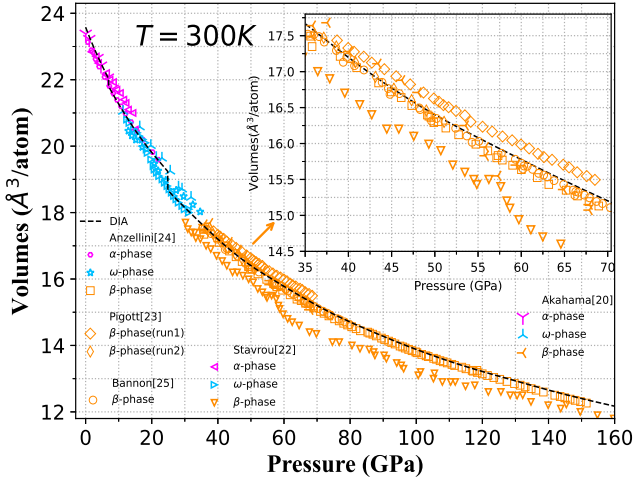}
  \caption{\label{fig:5}(Color Online) The $P$-$V$ EOS of pure Zr metal determined by DIA (black dashed line) up to 160 GPa along with those from experiments, where the $\alpha$-, $\omega$- and $\beta$-phases are colored in magenta, skyblue and yellow respectively and the inset exhibits details of the $\beta$-EOSs in a pressure range of $35$-$70$ GPa.}
\end{figure}

\subsection{The Stability of $\beta$-Phase at $300$K}
\label{sec:3:3}
By the computations of DIA,
the $\beta$-structure becomes the dominant phase with the minimum Gibbs FE at $300$K as the pressure being larger than $24.83$ GPa
and the calculated $P$-$V$ EOS of Zr up to $160$ GPa by Eq.(\ref{eq:2}) is shown in Fig.\ref{fig:5}.
The EOS of the $\beta$-phase does not show any discontinuity and is in excellent agreement with the three latest experimental observations with
the relative difference being $0.07\%$ in terms of the second run of Ref.\cite{pigott2019},
$0.42\%$ of Ref.\cite{bannon2021} and $0.62\%$ of Ref.\cite{anzellini2020},
while deviates about $1.2\%$ from the first run of Ref.\cite{pigott2019}.
Compared with the two experiments claiming the isostructural transition,
the relative differences are $0.96\%$ in terms of Ref.\cite{akahama1991},
and $4.15\%$ of Ref.\cite{stavrou2018},
which is by far the largest deviation of $P$-$V$ EOS between our computations and all the experiments.

By conducting the QMD simulations in Ref.\cite{stavrou2018},
Stavrou et al. attributed the observed anomaly to the anharmonicity of the $\beta$-phase,
which was confirmed by previous theoretical works at high-temperature zone (over $1000$K)\cite{zranharmon2}
and experimentally proved to be in close relationship to the strong glass forming ability of Zr with pressure over $6$ GPa and temperature at $900$K\cite{zhang2005},
though not reported at room-temperature conditions in any literatures.
As brought out by Anzellini et al.\cite{anzellini2020}, 
one questionable point towards the simulation lies in the usage of only one single $k$-point $(\frac{1}{8},\frac{1}{8},\frac{1}{8})$
that may lead to less precision of \emph{ab initio} interactomic interactions and
result in misleading instability when compared with the outcomes from the reproduced QMD simulation in Ref.\cite{anzellini2020} employing a $2\times2\times2$ $k$-mesh.
Another point should be mentioned is the simulated cooling rate of $10^{14}$ K/s in Ref.\cite{stavrou2018},
which is much faster than the critical cooling rate of $10$-$10^3$ K/s of Zr-based alloys\cite{bmg1,bmg2} or the estimated $10^{11}$-$10^{13}$ K/s of pure Zr metal\cite{bmg3,bmg4}.
With such a high rate,
the cooled down structure in the simulation would be more inclined to stay in a meta stable amorphous structure instead of a crystalline $\beta$-lattice.

In the model of DIA,
the thermal contributions from both harmonic and anharmonic lattice motions are naturally included and entangled together\cite{glc2},
which are on the whole reflected by the value of effect length defined in Eq.(\ref{eq:5})
that is decided by the shape of potential-energy curve $U'$ and the Boltzmann factor $e^{-\beta U'}$ at given temperatures.
Although we cannot separate these two parts,
the total thermal contributions,
consisting of kinetic energy of thermal atoms, entropy and additional work from thermal pressure,
can be quantitatively computed by subtracting the enthalpy at $0$K from the Gibbs FE at $300$K, $\mathcal G-\mathcal H$, and the result is shown in Fig.{\ref{fig:6}}.
As we can see,
the thermal contributions are at least over $120$ meV and getting larger from lower pressure zone ($\alpha$-phase) to higher pressure zone ($\beta$-phase) with a difference of about $60$-$80$ meV/atom.
Moreover, the thermal contributions abruptly become large at the two phase boundaries, especially for the $\omega\rightarrow\beta$,
indicating nontrivial impacts on the phase stabilities of Zr metal under high pressures.
With the above the comparisons and analysis,
we support the conclusion from latest experiments that there is no isostructural phase transitions in the $\beta$-phase.

\section{Conclusions}
\label{sec:4}
In conclusion,
by the criterion of the Gibbs FEs derived directly from PF,
the transition pressure of $\alpha\rightarrow\omega$ is determined to be $6.93$ GPa,
which is so far the only theoretical result coinciding well with experimental observations,
and the one of $\omega\rightarrow\beta$ is $24.83$ GPa,
which needs to be further validated in future experiments that, as we suggest,
may particularly pay attentions to the possible phase coexistence.
Additionally,
in considerations to the excellent agreement within $0.7\%$ difference between our obtained EOS in the high pressure zone up to $160$ GPa and those from latest experiments,
we are in favor of the argument that there does not exit isostructural phase transitions in the $\beta$-phase.
With quantitative comparisons with the results on the basis of enthalpy at $0$K,
we show that the thermal contributions are not negligible but are indeed important for the phase stability of Zr.
With the method of DIA and more advanced \emph{ab initio} computations,
the ensemble theory can be practically applied to investigate the phase behaviors and thermodynamic properties of more complex systems under extreme conditions.

\section{acknowledgement}
BYN is grateful to A.Dewaele for kindly providing the raw experimental data in Ref.\cite{anzellini2020}.
Part of the computational tasks was conducted in HPC platform supported by The Major Science and Technology Infrastructure Project of Material Genome Big-science Facilities Platform supported by Municipal Development and Reform Commission of Shenzhen.


%

\end{document}